\documentclass[journal=jctcce,manuscript=article]{achemso}

\usepackage[version=3]{mhchem} 
\usepackage{mathrsfs}
\usepackage{amsmath}
\usepackage{graphicx}
\usepackage{dcolumn}
\usepackage{bm}
\usepackage{bbm}
\usepackage{hyperref}
\usepackage{braket}
\usepackage{epstopdf}
\AppendGraphicsExtensions{.tiff}

\title{Ab initio calculations of quantum light-matter interactions in general electromagnetic environments}
\author{Mark Kamper Svendsen}
\email{mark-kamper.svendsen@mpsd.mpg.de}
\affiliation{Max Planck Institute for the Structure and Dynamics of Matter and
Center for Free-Electron Laser Science \& Department of Physics,
Luruper Chaussee 149, 22761 Hamburg, Germany}
\alsoaffiliation{Computational Atomic scale Materials Design~(CAMD), Department of Physics, Technical University of Denmark, 2800 Kgs. Lyngby,
Denmark}
\alsoaffiliation{Center for Computational Quantum Physics, Flatiron Institute, New York, New York 10010, USA}
\author{Kristian Sommer Thygesen}
\affiliation{Computational Atomic scale Materials Design~(CAMD), Department of Physics, Technical University of Denmark, 2800 Kgs. Lyngby,
Denmark}
\author{Angel Rubio}
\affiliation{Max Planck Institute for the Structure and Dynamics of Matter and
Center for Free-Electron Laser Science \& Department of Physics,
Luruper Chaussee 149, 22761 Hamburg, Germany}
\alsoaffiliation{Center for Computational Quantum Physics, Flatiron Institute, New York, New York 10010, USA}
\alsoaffiliation{Nano-Bio Spectroscopy Group and European Theoretical Spectroscopy Facility (ETSF), Universidad del Pa\'is Vasco (UPV/EHU), Av. Tolosa 72, 20018 San Sebastian, Spain}
\author{Johannes Flick}
\affiliation{Department of Physics, City College of New York, New York, New York 10031, USA}
\alsoaffiliation{Center for Computational Quantum Physics, Flatiron Institute, New York, New York 10010, USA}
\alsoaffiliation{Department of Physics, The Graduate Center, City University of New York, New York, New York 10016, USA}

\abbreviations{IR,NMR,UV}

\begin{document}


\begin{abstract}
The emerging field of strongly coupled light-matter systems has drawn significant attention in recent years because of the prospect of altering both physical and chemical properties of molecules and materials. Because this emerging field draws on ideas from both condensed-matter physics and quantum optics, it has attracted attention from theoreticians from both fields. While the former often employ accurate descriptions of the electronic structure of the systems the description of the electromagnetic environment is often oversimplified. In contrast, the latter often employs sophisticated descriptions of the electromagnetic environment, while using oversimplified few-level approximations of the electronic structure. Both approaches are problematic because the oversimplified descriptions of the electronic system are incapable of describing effects such as light-induced structural changes in the electronic system, while the oversimplified descriptions of the electromagnetic environments can lead to unphysical predictions because the light-matter interactions strengths are misrepresented. 

In this work, we overcome these shortcomings and present the first method which can quantitatively describe both the electronic system and general electromagnetic environments from first principles. We realize this by combining macroscopic QED (MQED) with Quantum Electrodynamical Density-functional Theory.
To exemplify this approach, we consider the example of an absorbing spherical cavity and study the impact of different parameters of both the environment and the electronic system on the transition from weak-to-strong coupling for different aromatic molecules. As part of this work, we also provide an easy-to-use tool to calculate the cavity coupling strengths for simple cavity setups. Our work is a significant step towards parameter-free \textit{ab initio} calculations for strongly coupled quantum light-matter systems and will help bridge the gap between theoretical methods and experiments in the field.
\end{abstract}
\section{Introduction}
When quantum light and matter interact strongly, hybrid light-matter polariton states emerge. These polaritonic states inherit properties from both light and matter, and it is therefore possible to alter the properties of either constituent by manipulating the other~\cite{basov2020polariton}. This flexibility opens the door to engineering both chemical and physical properties of matter with quantum light and has attracted significant attention in recent years~\cite{hutchison2012modifying,thomas2016ground,flick2017atoms,flick2018strong,hubener2021engineering,latini2021ferroelectric,appugliese2022breakdown}.

Polariton formation can be realized in resonant electromagnetic environments, the paradigmatic example of which is a Fabry-Perot cavity. In some setups, it is possible to enhance the vacuum fluctuations of the electromagnetic field sufficiently to drive the formation of polaritons even in the absence of actual light in the cavity. 
To realize such large light-matter coupling strengths it is, however, necessary to go beyond the paradigmatic cavity setup, consisting of two parallel mirrors, and often requires intricate meta- and nano-optical setups. Examples include plasmonic nanocavities~\cite{benz20216}, optical cavities using four-wave mixing schemes
~\cite{Pscherer2021}, metasurface systems~\cite{do2022}, self-assembled Casimir microcavities~\cite{Munkhbat2021}, deep-strong coupling in plasmonic nanoparticles~\cite{Mueller_2020}, and many more~\cite{basov2020polariton,simpkins2023control}. These cavity setups will generally feature complicated mode structures, especially with losses present, and therefore require descriptions beyond simple single mode descriptions.

Because this emerging field lives at the interface between condensed matter physics and quantum optics, a proper theoretical treatment simultaneously requires a quantitative description of the both electronic system and the quantum electromagnetic environment. However, to date, no method can properly account for both sides of the problem. Existing methods either apply few-level approximations to the electronic system, or oversimplified descriptions of the electromagnetic environment essentially treating the light-matter coupling strengths as free parameters.
This is problematic because the light-matter interaction is very sensitive to both phase matching and energetic alignment of field and matter, as well as the spatial overlap of the electronic wave functions and the modes of the electromagnetic field~\cite{svendsen2021combining}. Furthermore, simplified descriptions of the electronic system are incapable of describing effects such as light-induced structural changes in the electronic system and are generally not applicable in the ultra- and deep strong coupling regimes. Finally, the oversimplified descriptions of the electromagnetic environments can lead to unphysical predictions because the light-matter interaction strengths are misrepresented~\cite{svendsen2021combining}.


In this paper we present the first method that accounts quantitatively for both the electromagnetic environment and the full electronic structure of matter by combining Macroscopic Quantum Electrodynamics~(MQED)~\cite{scheel2009macroscopic} with Quantum Electrodynamical Density-functional Theory~(QEDFT)~\cite{ruggenthaler2014quantum}. We have previously shown how standard DFT can be combined with MQED to provide a quantitative, first principles description of the quantum light-matter interactions for real cavity setups in the weak and intermediate coupling regimes beyond the dipole approximation~\cite{svendsen2021combining}. However, the previous work relied on a wave-function ansatz which only considered a subset of the electronic structure. This is expected to become problematic in the ultra- and deep strong coupling regimes. Furthermore, our previous method would also fail capture light-induced structural changes in the electronic system. In this work, we overcome these limitations and present a general method that is applicable in all regimes of light-matter coupling. Our new methodology allows us to study the interaction of the full electronic structure of electronic systems with realistic electromagnetic environments. We exemplify our approach on a spherical micro-cavity and highlight that the intricate interplay between the cavity geometry and material composition, and the electronic structure of the molecules has a profound impact on the light-matter coupling and the transition from weak to strong coupling. These results highlight the need for a quantitative description of both the electromagnetic environment and molecular system. This work is a significant step towards parameter free \textit{ab initio} calculations for strongly coupled quantum light-matter systems and will help bridge the gap between theoretical methods and experiments in the field.

\section{Theory and Methodology}

\subsection{Macroscopic Quantum Electrodynamics}

Macroscopic Quantum Electrodynamics (MQED) is a framework for quantizing the electromagnetic field in the presence of arbitrary absorbing or dispersing environments~\cite{scheel2009macroscopic, feist2021macroscopic,svendsen2021combining}. The central object in MQED is the \textit{classical} Green's function that solves the Helmholtz equation for a point source, the so-called dyadic Green's function (DGF)\cite{chew1990waves},
\begin{align}\label{eq:def_DGF}
    \left[\boldsymbol{\nabla}\times\kappa(\boldsymbol{r},\omega) \boldsymbol{\nabla}\times -\frac{\omega^2}{c^2}\epsilon(\boldsymbol{r},\omega)\right]\boldsymbol{G}(\boldsymbol{r},\boldsymbol{r'}, \omega) = \boldsymbol{I}\delta(\boldsymbol{r}-\boldsymbol{r'}).
\end{align}
Here $c$ is the speed of light, $\omega$ is the angular frequency, $\epsilon(\boldsymbol{r},\omega)$ and $\kappa(\boldsymbol{r},\omega)=\mu^{-1}(\boldsymbol{r},\omega)$ are the spatially dependent dielectric function and inverse magnetic permability respectively, and $\boldsymbol{I}$ is the unit dyad. The DGF is of central importance to the quantized theory of electromagnetic fields in lossy environments because it simultaneously carries the information about the electromagnetic boundary conditions, and serves as a projector from the coupled light-matter system onto the electromagnetic degrees of freedom~\cite{scheel2009macroscopic}. 

For a spatially local magnetoelectric medium in the nonrelativistic limit, the MQED expansion of the electric field in the Power-Zienau-Woolley (PZW) frame~\cite{woolley1980, babiker1983} (multipolar gauge) can be written as~\cite{scheel2009macroscopic,feist2021macroscopic},
\begin{align}
\label{eq:efield}
    \boldsymbol{\hat{E}}(\boldsymbol r) &= \int d\omega \boldsymbol{\hat{E}}(\boldsymbol r, \omega) + \mathrm{h.c.}, \\
    \boldsymbol{\hat{E}}(\boldsymbol r, \omega) &= \sum_{\lambda = e,m}\int d^3r' \boldsymbol{G}_\lambda(\boldsymbol r, \boldsymbol r', \omega)\cdot \boldsymbol{\hat{f}}_\lambda(\boldsymbol r', \omega).
\end{align}
Here $\boldsymbol{\hat{f}}_\lambda(\boldsymbol r', \omega)$ are the spatially resolved polaritonic field operators of MQED which fulfill the commutation relations of the quantum harmonic oscillator, and $\boldsymbol{G}_e(\boldsymbol r, \boldsymbol r', \omega)$ and  $\boldsymbol{G}_m(\boldsymbol r, \boldsymbol r', \omega)$ are the electric and magnetic components of the DGF respectively,
\begin{align}
    &\boldsymbol{G}_e(\boldsymbol r, \boldsymbol r', \omega) = i\frac{\omega^2}{c^2}\sqrt{\frac{\hbar}{\pi\epsilon_0}\mathrm{Im}\epsilon(\boldsymbol{r},\omega)}\boldsymbol{G}(\boldsymbol r, \boldsymbol r', \omega), \\
    &\boldsymbol{G}_m(\boldsymbol r, \boldsymbol r', \omega) = -i\frac{\omega}{c}\sqrt{\frac{\hbar}{\pi\epsilon_0}\mathrm{Im}\mu(\boldsymbol{r},\omega)}\boldsymbol{G}(\boldsymbol r, \boldsymbol r', \omega)\times\boldsymbol{\nabla'}.
\end{align}

In the following we neglect magnetic interactions and consider the coupling between light and matter within the dipole approximation. Therefore, if we consider a set of emitters $i$ with positions (centers of charge) $\boldsymbol r_i$, the interaction only samples the electromagnetic field at these positions. In this sense, the full electric field $\hat{\boldsymbol E}(\boldsymbol r)$ in Eq.~\ref{eq:efield}  contains more information than strictly necessary to describe the light-matter interaction completely. As discussed in Refs \cite{feist2021macroscopic,Buhmann2008,Hummer2013}, it is therefore possible to arrive at a significantly more compact expression by alternatively expanding the electric field in terms of a set of $\mathcal{N}$ explicitly orthogonalized bright modes at each frequency,
\begin{align}
\label{eq:efield-bright}
    &\boldsymbol{\hat{E}}(\boldsymbol r) = \sum_{i=1}^{\mathcal{N}} \int_0^\infty d\omega \boldsymbol{E}_i(\boldsymbol r, \omega)\hat{C}_i(\omega) +\mathrm{H.c.},\\
    & \boldsymbol{E}_i(\boldsymbol r, \omega) = \frac{\hbar\omega^2}{\pi\epsilon_0c^2}\sum_{j=1}^{\mathcal{N}} V^*_{ij}(\omega)\frac{\mathrm{Im} \boldsymbol G(\boldsymbol r, \boldsymbol r_j, \omega)\cdot\boldsymbol{n}_j}{G_j(\omega)}.\label{eq:e_field_dgf}
\end{align}
Here $\hat{C}^{(\dagger)}_i(\omega)$ destroys (creates) a photon in the $i$th bright mode. $\boldsymbol{E}_i(\boldsymbol r, \omega)$ describes the spatial mode function of the electric field associated with mode $i$. The normalisation factor $G_j(\omega)$ is the square root of the dipole spectral density\cite{novotny2012principles},
\begin{align}
    G_j(\omega) = \left(\frac{\hbar\omega^2}{\pi\epsilon_0c^2}\boldsymbol{n}_j\cdot\mathrm{Im}\boldsymbol{G}(\boldsymbol r_j, \boldsymbol r_j, \omega)\cdot\boldsymbol{n}_j\right)^{1/2}.
\end{align}
Finally, the matrix $V_{ij}(\omega)$  is the transformation matrix which obeys $\boldsymbol{V}(\omega)\boldsymbol{S}(\omega)\boldsymbol{V}^\dagger(\omega) = \boldsymbol{I}$ where 
\begin{align}
    S_{ij}(\omega) =  \frac{\hbar\omega^2}{\pi\epsilon_0c^2}\frac{\boldsymbol{n}_i\cdot\mathrm{Im}\boldsymbol{G}(\boldsymbol r_i, \boldsymbol r_j, \omega)\cdot\boldsymbol{n}_j}{G_i(\omega)G_j(\omega)},
\end{align}
and it is a result of the mode-orthogonalization inherent to the emitter-centered representation~\cite{feist2021macroscopic}. The number of emitter centered modes per frequency, $\mathcal{N}$, is equal to the number of emitter positions times the number of dipole orientations considered. In this work, we consider a single emitter position $\boldsymbol{r}_0$ and the full three-dimensional space of dipole orientations. This results in three emitter-centered modes per frequency. Importantly, while the bright mode representation of the MQED field is able to account for the spatial dependence of the inter-emitter interaction, the \emph{local} coupling of each emitter to the field is described within the dipole approximation. Even though the electric field in Eq.~\ref{eq:efield-bright} retains spatial dependence it is therefore not a fully beyond dipole approximation representation of the MQED field. We note that the description of the local coupling within the dipole approximation is in principle problematic for the high frequency modes which are inherently included in Eq.~\ref{eq:efield-bright}. Because no real electronic system is truly point-like, and the field expansion in principle includes modes of arbitrarily large frequencies, the coupling of the emitters to some of these modes would inherently require a beyond dipole approximation. However, as discussed further below, it is necessary in practice to truncate the mode expansion at some upper frequency, and for the modes included in the actual calculations the dipole approximation is justified. For a discussion of how beyond dipole approximation light-matter coupling can be explored within the MQED framework with reduced models of the electronic structure, we refer to our previous work in Ref.~\cite{svendsen2021combining}.

We want to express the total Hamiltonian of the coupled light-matter system in a similar form as used previously in Refs.~\cite{tokatly2013time,flick2019light,wang2021light}. Therefore we write (see Supplementary Note \ref{app:PF-in-terms-of-emitter-centered-modes}),
\begin{align}
\label{eq:hamiltonian}
    \mathcal{H} =& \mathcal{H}_\mathrm{Mat} + \nonumber \\ 
    &\frac{1}{2}\sum_{i=1}^{\mathcal{N}}\int_0^\infty d\omega \left\{\hat{p}_i(\omega)^2 + \omega^2\left[\hat{q}_i(\omega) + \frac{\boldsymbol \lambda_i(\omega)}{\omega}\cdot{\hat{\boldsymbol R}}\right]^2\right\},
\end{align}
where $\mathcal{H}_\mathrm{Mat}$ is the standard Coulomb gauge matter Hamiltonian describing the electronic system, and $\boldsymbol \lambda_i(\omega)$ is the cavity field strength of the $i$'th bright mode at frequency $\omega$,
\begin{align}
\label{eq:lambda}
\boldsymbol \lambda_i(\omega)= e\left(\frac{2}{\hbar\omega}\right)^{1/2} \boldsymbol{E}_i(\boldsymbol r_0,\omega),
\end{align}
where $\boldsymbol r_0$ denotes the electronic center of charge and $\boldsymbol{E}_i(\boldsymbol r_0,\omega)$ is defined in terms of the dyadic Greens function in Eq.~\ref{eq:e_field_dgf}. The cavity field strengths of an arbitrary electromagnetic environment are thus fully determined by the DGF. We have further introduced the photon field quantities $\hat p_i$ and $\hat q_i$ that are connected to the magnetic- and electric field in their corresponding mode and are given explicitly by $\hat{q}_i(\omega) = \left(\frac{\hbar}{2\omega}\right)^{1/2}(\hat{C}_i(\omega) + \hat{C}^\dagger_i(\omega))$ and $\hat{p}_i(\omega) = \left(\frac{\hbar\omega}{2}\right)^{1/2}(\hat{C}_i(\omega) - \hat{C}^\dagger_i(\omega))$. In terms of these new quantities, the electric field expansion at the center of charge $\boldsymbol r_0$ reads,
\begin{align}\label{eq:results:qedft-mqed:emitter-c-E-in-terms-of-lambda}
    &\boldsymbol{\hat{E}}(\boldsymbol r_0) = \sum_{i=1}^{\mathcal{N}} \int_0^\infty  \frac{\omega}{e}\boldsymbol{\lambda}_i(\omega)\hat{q}_i(\omega)d\omega.
\end{align}
The light-matter interaction further contains the dipole moment operator for $N_e$ electrons with position $\hat{\boldsymbol{r}}_i$, ${\hat{\boldsymbol R}=\sum_{i=1}^{N_e} \hat{\boldsymbol{r}}_i}$, which gives rise to an explicit electron-photon interaction and the dipole-self energy term ( $\sim\left(\boldsymbol\lambda_i(\omega) \cdot \boldsymbol{\hat R}\right)^2$). While the dipole self-energy term is often neglected or absorbed into the matter Hamiltonian in the context of MQED~\cite{scheel2009macroscopic,buhmann2013dispersion,feist2021macroscopic}, recent works have shown that its inclusion is critical to ensure gauge invariance, and the stability of the coupled light-matter system in the strong- and ultra strong coupling regimes~\cite{rokaj2018light,schafer2020relevance}.When included explicitly, the dipole self-energy term is most commonly expanded in terms of the modes of the electromagnetic field~\cite{tokatly2013time,ruggenthaler2014quantum}. As discussed in Supplementary Note A, we follow the same procedure in this work to arrive at the dipole self-energy term in Eq.~\ref{eq:hamiltonian}. In practice, most numerical implementations apply some truncation of the photonic Hilbert space. This is also true for our framework as discussed in the following sections. It has been shown that any truncation of the photonic Hilbert space must be accompanied by a consistent truncation of the dipole self-energy term to avoid unphysical predictions of both ground and excited state properties~\cite{taylor2022resolving}. In this work we therefore include the same modes in the expansion of the light-matter interaction term and the dipole self-energy term.

The interaction of the electronic system with the electric field within the dipole approximation can thus be expressed as the interaction of the electronic system and a continuous set of quantum harmonic oscillator modes via the dipole moment of the electronic system. We thus arrive at a Hamiltonian which is able to describe the full electronic structure of matter in the presence of a realistic electromagnetic environment. While the electronic system is described fully \textit{ab initio}, it is important to note that within MQED, the environment is described via its spatially dependent dielectric properties. These can for example be calculated themselves using \textit{ab initio} methodology for the materials making up the cavity structure~\cite{onida2002electronic,yan2011linear,andersen2015dielectric,svendsen2022computational}, or be described using simpler models of dielectric response such as e.g. the Drude model or the Lorentz Oscillator model~\cite{saleh2019fundamentals,novotny2012principles}. Note that our formulation addresses both the problem of how to formulate the length gauge Hamiltonian in the presence of optical losses, and allows for the explicit calculation of the cavity field strengths in terms of the boundary conditions set by the cavity via the DGF of the electromagnetic field.

\subsection{Quantum Electrodynamical Density Functional Theory}
QEDFT is a generalization of density functional theory (DFT) for electronic systems interacting strongly with quantized modes of the electromagnetic field~\cite{ruggenthaler2014quantum}. The method of QEDFT can describe the full electronic structure of the matter as well as the interaction of electrons with photons, and represents a good compromise between accuracy and computational cost. QEDFT has been successfully applied to describe both the ground-state~\cite{pellegrini2015optimized,flick2018ab}, and excited-states~\cite{flick2019light,yang2021quantum} of single (few) molecules strongly coupled to quantized modes of light, as well as for applications in polaritonic chemistry~\cite{Sch_fer_2022}. The existence of the QEDFT formulation can be proven under very general conditions~\cite{ruggenthaler2014quantum} and it in principle allows for the treatment of coupled light-matter systems with many electrons and many photonic modes under very general conditions.  However, currently, most practical implementations of QEDFT are based on the dipole approximation for the light-matter coupling and a discrete mode expansion of the electromagnetic field. The latter implies that the material or molecular system of interest is embedded in a lossless electromagnetic medium. There have been previous studies applying the QEDFT formulation to lossy optical cavities to describe e.g. photon losses through cavity mirrors  \cite{flick2019light,wang2021light,kudlis2021dissipation}. Here, different models of the optical cavity were used, but no general connection between the cavity field parameters and the optical environment for absorbing and dispersing magnetoelectric bodies has yet been established. As a result, the electron-photon coupling parameters, while in principle connected to the physical quantity of the vacuum electric field at the center of charge of the system, are then in practice often treated as free parameters. This highlights the other current practical limitation of QEDFT: Treating the cavity coupling parameters as essentially free parameters make quantitative calculations and direct comparison with experiments hard. Here we overcome these practical limitations by representing the electromagnetic environment in terms of the MQED field expansion which allows us to accurately account for both optical losses and the quantitative magnitude of the cavity field strengths. The result is a parameter-free description of arbitrary electronic systems coupled to general, lossy electromagnetic environments.

To calculate the excited-state properties of the coupled electron-photon problem defined by the Hamiltonian in Eq.~\ref{eq:hamiltonian},
we employ the linear-response formulation of QEDFT, which has been described in detail in Refs.~\cite{flick2019light, welahkuh2022, yang2021quantum}. Importantly, this method considers the \emph{full} electronic structure of the matter system and goes beyond the rotating wave approximation. We specifically employ the generalized Casida formulation of linear response QEDFT to calculate the oscillator strengths of the many-body excitations of the coupled light-matter (polaritonic) system. The generalized Casida equation reads,
\begin{align}\label{eq:theory-lr:generalized-casida}
    \begin{bmatrix}
    U & V\\
    V^T & \omega_\alpha 
    \end{bmatrix} 
    \begin{bmatrix}
    \boldsymbol{F} \\
    \boldsymbol{P}  
    \end{bmatrix}= \Omega^2_S
    \begin{bmatrix}
    \boldsymbol{F} \\
    \boldsymbol{P}  
    \end{bmatrix}.
\end{align}
Here $\omega_\alpha$ is a diagonal matrix with the frequency of photon modes in the diagonal, $U$ accounts for the coupling between the electrons and $V$ the coupling between electrons and photons. The eigenvalues $\Omega_S^2$ of the Casida equation are the square of the electron-photon excitation energies $\omega_I$. $\boldsymbol{F}$ and $\boldsymbol{P}$ describe the Casida eigenvectors of the matter and photon block, respectively. The square norm of the two gives the electronic- and photonic fraction of the excitation respectively. Introducing the pair orbital index $S = (ia)$, corresponding to the pair orbital $\Phi_S(\boldsymbol{r}) = \phi_i(\boldsymbol{r})\phi_a^*(\boldsymbol{r})$ with energy $\epsilon_S = \epsilon_a - \epsilon_i$, $U$ can be expanded as,
\begin{align}
    &U_{SS'}(\Omega_s) = \epsilon_S^2\delta_{SS'} + 2\epsilon_S^{1/2}K_{SS'}(\Omega_S)\epsilon_{S'}^{1/2},\\
    &K_{SS'}(\Omega_S) = \int \int d\boldsymbol{r}d\boldsymbol{r'}\Phi_S(\boldsymbol{r})f^n_\mathrm{Mxc}(\boldsymbol{r},\boldsymbol{r'},\Omega_S)\Phi_{S'}(\boldsymbol{r'}).
\end{align}
If $N_\mathrm{pair}$ pair-orbitals are included in the calculation, $U$ is therefore an $N_\mathrm{pair}\times N_\mathrm{pair}$ matrix. $V_{\alpha S}$ is the matrix accounting for the coupling between the electrons and the photon modes,
\begin{align}
    &V_{\alpha S}(\Omega_S) = 2\sqrt{\epsilon_S M_{\alpha S}(\Omega_S)N_{\alpha S}\omega_\alpha},\\
    &M_{\alpha,S}(\Omega_S) = \int d\boldsymbol{r}\Phi_{S}(\boldsymbol{r})f_\mathrm{Mxc}^{q_\alpha}(\boldsymbol{r}, \Omega_S), \\
    &N_{\alpha, S} = \frac{1}{2\omega_\alpha^2}\int d\boldsymbol{r}\Phi_{S}(\boldsymbol{r})g_\mathrm{M}^{n_\alpha}(\boldsymbol{r}).
\end{align}
The size of these matrices will be $ N_\mathrm{modes}\times N_\mathrm{pair}$. We note that these equations include the exchange correlation kernels $f_{Mxc}^n$, $f_{Mxc}^q$ and $g_\mathrm{M}^{n_\alpha}$ which have to be approximated in practice. So far, the available exchange-correlation functionals for QEDFT in the time-domain are still limited~\cite{ pellegrini2015optimized,flick2018ab}, but recent developments for efficient density functionals based on the photon-free formulation of QEDFT~\cite{schafer2021making} or the QEDFT fluctuation-dissipation theorem~\cite{flick2021simple} are promising. Finally, note that when neglecting the quantized light, only the top left block in Eq. \ref{eq:theory-lr:generalized-casida} survives and one recovers the standard Casida formulation of linear response TDDFT for finite systems~\cite{casida1995time}. 

In general, Eq.~\ref{eq:theory-lr:generalized-casida} requires a self-consistent solution as $U$ and $V$ depend on the eigenvalues $\Omega_S$. In this work, we neglect any exchange-correlation contribution in the photonic exchange-correlation kernels and apply the mean-field photonic random phase approximation. We refer to Ref.~\cite{flick2019light} for a thorough discussion of this approximation. In this case, the connection between the photonic exchange-correlation kernels and the cavity field strengths in mode $\alpha$ is $f_\mathrm{Mxc}^{q_\alpha} =-\omega_\alpha \boldsymbol\lambda_\alpha\cdot e\textbf{r}$ and $g_\mathrm{M}^{n_\alpha}= -\omega^2_\alpha \boldsymbol\lambda_\alpha\cdot e\textbf{r}$. We emphasize here the connection of the linear-response exchange-correlation kernels of QEDFT with the dyadic Greens function of the MQED framework via the cavity field strength $\boldsymbol \lambda_\alpha$ defined in Eqs. ~\ref{eq:e_field_dgf} and \ref{eq:lambda}. Importantly, the Casida formulation works for a discrete set of cavity field strengths. In practice, it is therefore necessary to employ a dense, discretized sampling of the continuous frequency expressions for the coupling strengths. As discussed in Supplementary Note C, we employ uniform sampling in this work. It is worth noting that recently more efficient sampling methods that result in a lower number of modes have been put forth in the literature for master equation based approaches~\cite{franke2019quantization,medina2021few,sanchez2021few}. While it is not immediately clear that these approaches can be integrated with the framework presented here because the quasi-normal modes are made interacting in these schemes, it would be an interesting avenue of future research to investigate whether such sampling methods can be used in our formalism. 




As discussed in Ref. \cite{flick2019light} it is possible to calculate the oscillator strengths of the coupled system in terms of the eigenvectors and eigenvalues of Eq.~\ref{eq:theory-lr:generalized-casida} and these fully describe the linear response of the coupled light-matter system. In this work, we characterize the response of the system using the oscillator strengths of the electronic polarizability, $f_I$. For the transition between the many body ground state $\Psi_0$ and the excited state $\Psi_I$ the oscillator strength is formally given in terms of the transition dipole matrix element as $f_I = \frac{2}{3}\omega_I\sum_{i=1}^3|\bra{\Psi_0}e r_i \ket{\Psi_I}|^2$. Importantly, we note that even in the presence of quantum light-matter interaction, these oscillator strengths still fulfill the $f$-sum rule $\sum_I f_I = N_e$~\cite{flick2019light}, where $N_e$ denotes the number of electrons. The excitation spectrum is then characterized by the strength function~\cite{flick2019light},
\begin{align}
    S(\omega) = \sum_I f_I\delta(\omega -\omega_I),
\end{align}
where the sum index $I$ runs over the many-body excitations of the coupled light-matter (polaritonic) system. We further emphasize that in this work we do not apply any broadening to the spectra and that the width of the peaks is determined completely by the electromagnetic environment. This highlights a further advantage of using the QEDFT-MQED combination, namely that this provides a natural description of the transition linewidths as they relate to decay induced by the electromagnetic environment. This is relevant beyond cavity QED settings and paves the way towards TDDFT without artificial linewidths.

\section{Results and discussion}
\subsection{Spherical microcavity}
\begin{figure}
    \centering
    \includegraphics[width=0.9\textwidth]{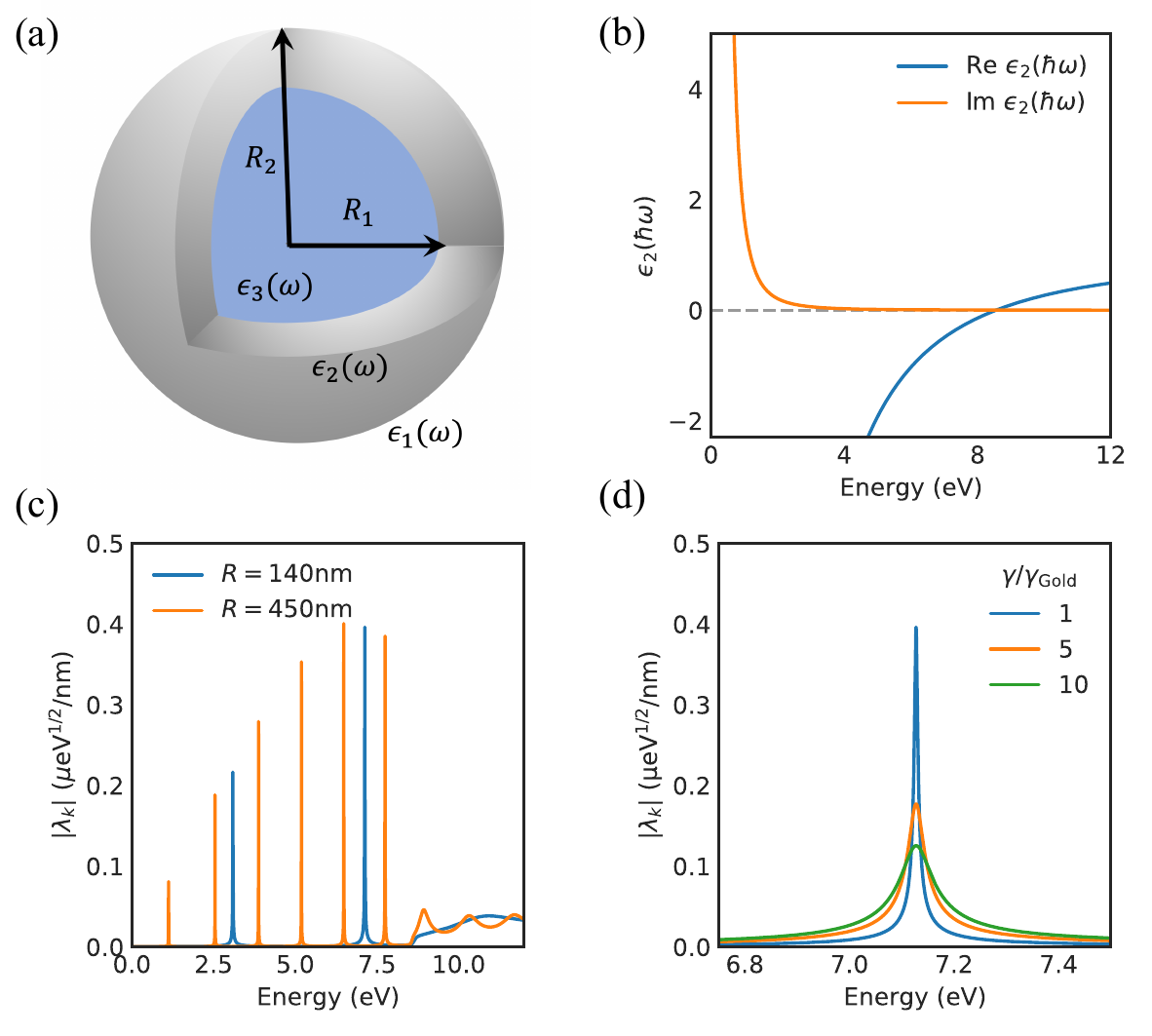}
    \caption{\textbf{Spherical microcavity setup:} (a) Illustration of the spherical microcavity setup. (b) The Drude model dielectric function of gold. (c) The modes of the spherical microcavity plotted for two different radii, $R=140$nm in blue and $R=450$nm in orange. The mode structure is shown with a sampling density of $10$ points/meV. (d) The impact from the Drude dampening parameter on the cavity resonances, illustrated by focusing on the mode around 7.1 eV in the cavity with $R=140$nm.}
    \label{supplementary-fig:gold-cavity-example}
\end{figure}

To exemplify the developed approach, we consider the spherically layered microcavity also considered in Refs. \cite{li1994electromagnetic,dung2000spontaneous}. As shown in Fig. \ref{supplementary-fig:gold-cavity-example}a, the spherical cavity consists of three concentric spherical layers, each characterized by a frequency dependent dielectric function, $\epsilon_n(\omega)$. 

Due to spherical symmetry of the cavity setup, the DGF is most efficiently expanded onto a set of vector spherical harmonics~\cite{chew1990waves}. For a general emitter position inside the cavity, it is necessary to carefully converge the number of vector spherical harmonics used in the calculation of the DGF. However, as we discuss in Supplementary Note B, if the emitter is placed in the center of the cavity, the situation simplifies significantly. In this case only the lowest order transverse magnetic mode of the cavity contributes and we can write~\cite{dung2000spontaneous},
\begin{align}
    \mathrm{Im}\boldsymbol{G}(\boldsymbol 0, \boldsymbol 0,\omega) = \frac{\omega}{6\pi c}\left[1 + \mathrm{Re} \left(r^\mathrm{TM}_{n=1}(\omega)\right)\right]\boldsymbol{I},
\end{align}
where $r^\mathrm{TM}_{n=1}(\omega)$ is the reflection coefficient for the lowest order transverse magnetic~(TM) mode at the interface between the cavity region and the metal shell. $r^\mathrm{TM}_{n=1}(\omega)$ can be calculated by invoking the standard electromagnetic boundary conditions at the interfaces between the different regions of the cavity geometry. Notice that $\boldsymbol{n}_i\cdot\mathrm{Im}\boldsymbol{G}(\boldsymbol 0, \boldsymbol 0, \omega)\cdot\boldsymbol{n}_j \propto \delta_{ij}$ which means that the mode orthogonalization is trivial in this case, and the cavity field strengths can be derived directly using Eq.~\ref{eq:lambda},
\begin{align}
    \boldsymbol \lambda_i(\omega) = e \left(\frac{\omega^2}{3\pi^2\epsilon_0c^3}\left[1 + \mathrm{Re}\left(r^\mathrm{TM}_{n=1}(\omega)\right)\right]\right)^{1/2}\boldsymbol{n}_i.
\end{align}

\subsubsection{Drude metal shell}\label{sec:drude_metal_section}

We now consider the case where the inner and outer regions consist of vacuum. For the middle region we consider a simple but realistic model of a metallic mirror, namely a Drude metal with dielectric function,
\begin{align}\label{eq:drude-model}
    \epsilon_2(\omega) = 1 - \frac{\omega_p^2}{\omega^2 +i\gamma\omega},
\end{align}
where $\omega_p$ is the metal plasma frequency and $\gamma$ is the Drude dampening rate. As a concrete example of a metal, we use the Drude parameters for gold taken from Ref.~\cite{blaber2009search}. Within the Drude mode, the plasma frequency of gold is around 8.5 eV~\footnote{It should be noted that the Drude model is not a good model for the dielectric properties of gold due to the neglect of d-band absorption. Additionally, it should be mentioned that the Drude model will fail to account for size dependent resonance shifts and broadening resulting from the nonlocal response and surface enhanced Landau dampening in small metallic structures~\cite{mortensen2014generalized,svendsen2020role}.}. This results in the dielectric function shown in Figure. \ref{supplementary-fig:gold-cavity-example}b. Below the plasma frequency, the real part of the dielectric function will be negative and the material surface will consequently be highly reflective. Above the plasma frequency, the real part of the dielectric function becomes positive and the material will loose its metallic characteristics resulting in a significant loss of surface reflectivity. We note that because the $V$ matrices in Eq.~\ref{eq:theory-lr:generalized-casida} grows with the number of photon modes, it is necessary in practice to truncate the electromagnetic environment at some frequency. For the spherical cavity, $\omega_p$ provides a natural cutoff and we therefore include photon modes up to $\omega_p$ in the calculations. Further details of the QEDFT calculations are given in Supplementary Note D.

In Figure \ref{supplementary-fig:gold-cavity-example}c the mode structure of the cavity setup is shown for two different cavity radii, $R=140$nm and $R=450$nm. It is clearly observed that the number of modes, as well as their spectral position, is directly linked to the radius of the microcavity. Furthermore, we clearly observe that above the gold plasma frequency, the mirrors lose their reflectivity which results in the loss of the sharp mode structure which is replaced by a continuum. This highlights that the formalism we present is able to directly link the cavity field strengths to the real cavity setup made of real materials. Figure \ref{supplementary-fig:gold-cavity-example}d zooms in on the mode around 7.1 eV in the cavity with $R=140$nm and shows the effect of changing the Drude dampening parameter. We clearly observe that the width of the cavity mode increases with increasing dampening in the metal which highlights the connection between the width of the cavity modes and the losses in the gold. 

This example highlights how the use of the emitter centered representation of MQED allows us to directly and uniquely relate the light-matter coupling strength to a real electromagnetic environment and connect it to the QEDFT formalism. While this is a relatively simple example, the approach is general and works analogously for an arbitrary electromagnetic environment provided that the DGF can be determined.

\subsubsection{Adding an emitter to the cavity}
\begin{figure*}
    \centering
    \includegraphics[width=\textwidth]{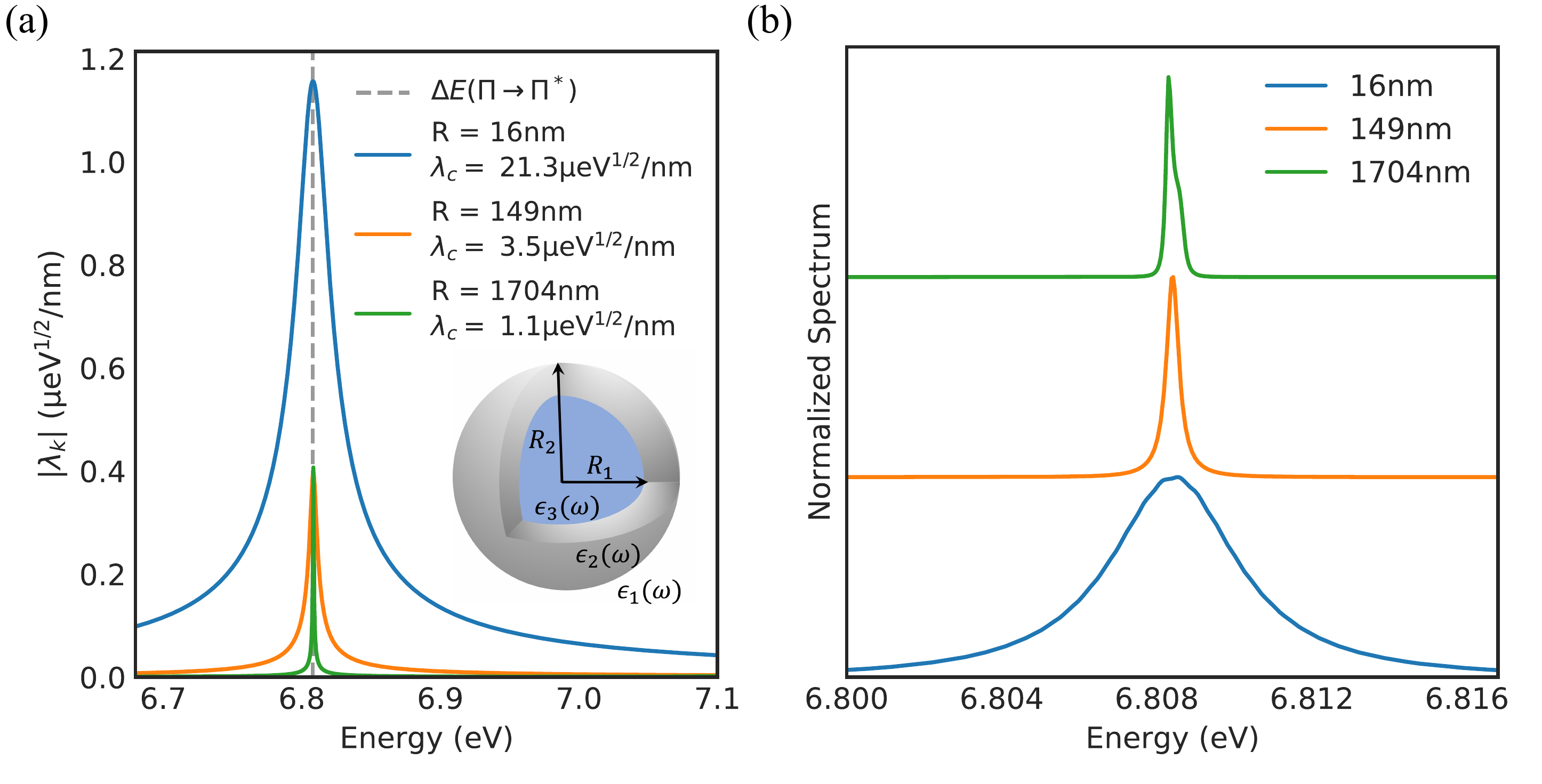}
    \caption{(a) Different radii of the spherical microcavity which host a resonance that aligns with the $\Pi\rightarrow\Pi^*$ of the benzene molecule. The total light-matter coupling strength of the modes is also shown in the legend, and it can be clearly observed that the cavity-coupling strength grows with decreasing cavity radius. The mode structure is shown with a sampling density of $10$ points/meV. (b) The linear absorption spectrum of benzene as a function of cavity radius showing a clear radius dependent Purcell enhancement.}
    \label{fig:cavity-illustration-figure}
\end{figure*}

We next add a benzene molecule to the cavity. Benzene is chosen mainly because of it prevalence as a test system in the existing TDDFT and QEDFT literature on strong coupling~\cite{flick2019light,wang2021light,rossi2019strong,flick2018ab}, but we emphasize that the method can treat arbitrary finite electronic systems. We focus on finding a cavity configurations with a mode resonant with the $\Pi\rightarrow\Pi^*$ transition of the benzene molecule. The first step is to determine the spectral position of this transition. Using the Casida linear response QEDFT framework without photons, we find that the transition occurs at 6.808 eV (182 nm) in free space and that the transition dipole moment in-plane is $|\boldsymbol{d}| = 0.096\, e\, \mathrm{nm}$. This is consistent with previous TDDFT calculations for benzene~\cite{flick2019light,wang2021light}. \\

As shown in Fig. \ref{fig:cavity-illustration-figure}a, it is possible to find different radii of the gold microcavity for which there is a cavity mode resonant with the benzene $\Pi\rightarrow\Pi^*$ electronic transition. As expected, the cavity field strength increases as the cavity is made smaller. All but the smallest cavity are optical cavities in the sense that the characteristic dimension of the cavity, the radius, is larger than half the wavelength of the transition. For the smallest cavity of radius 16 nm, a significant increase in the coupling strength relative to the other sizes is observed. This happens exactly because this cavity is sub-wavelength sized and therefore significant near-field coupling to the surface plasmon mode of the gold starts to occur. We mention in passing that while we solve the coupled system using QEDFT, the approach presented here for the calculation of cavity field strengths is applicable regardless of the method used to solve the coupled system. For completeness, we therefore note that in terms of the light-matter coupling strengths, $g$, more commonly used in quantum optics, the three cavities shown in Fig. \ref{fig:cavity-illustration-figure}a respectively correspond to $\hbar g = 3.81$ eV, $\hbar g = 0.62$ eV and $\hbar g = 0.21$ eV for the $\Pi\rightarrow\Pi^*$ transition of the benzene molecule. Importantly, these light-matter coupling strengths are a property of the coupled system and they are therefore specific to both the cavity and molecular transition under consideration. One should therefore be careful with general conclusions on the magnitude of light-matter interactions based on these numbers. We provide more discussion in Supplementary Note \ref{app:lm-coupling}. \newline 
Fig. \ref{fig:cavity-illustration-figure}b shows the linear absorption spectra of the coupled emitter-cavity system calculated for the different cavity radii using the linear response QEDFT method. We emphasize that all linewidths in the figure are \textit{true} linewidths in the sense that they are not related to any broadening parameters in the QEDFT calculation and only reflect the density of states in the optical environment. A radius-dependent Purcell enhancement with decreasing radius is clearly observed, reflecting the reduction in radiative lifetime resulting from the altered optical environment. However, it is not possible to achieve strong coupling with a single benzene molecule using the gold-shell cavity. We attribute this to the fact that as the coupling strength gets larger with decreasing radius, the optical losses also increase, resulting in a broader cavity resonance. We note in passing that the Purcell enhancement for the smallest $R = 16$ nm cavity is around 3500 which means that the local field enhancement at the center of the spherical microcavity is comparable to what is found in experiments with plasmonic microcavities~\cite{akselrod2014probing}. \\

\begin{figure*}[tb]
        \centering
        \includegraphics[width=\linewidth]{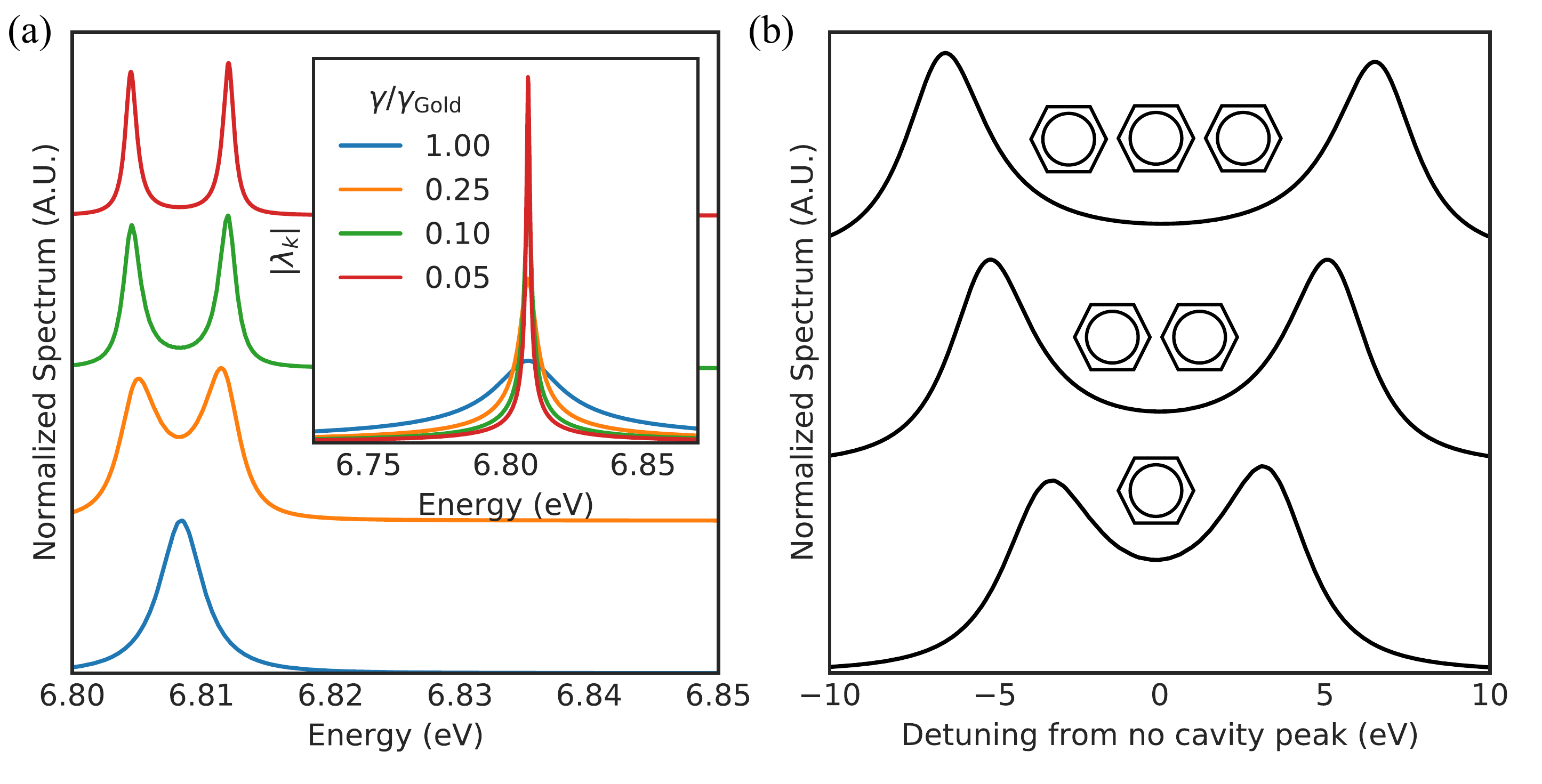}
        \caption{(a) The linear absorption spectrum of benzene in the gold cavity from Fig. \ref{fig:cavity-illustration-figure}a with an inner radius $R=16$nm as a function of the Drude dampening in the metallic mirror region. (b) The linear absorption spectrum in a cavity with with an inner radius $R=16$nm and Drude dampening $\gamma = \gamma_\mathrm{gold}/4$ as a function of the number of benzene molecules.}
        \label{fig:spherical_cavity_benzene}
\end{figure*}

The reason that it was impossible to reach strong coupling with benzene in the gold cavity was the losses of the cavity mode.  In an attempt to reach the SC regime, we therefore next seek to reduce the losses in the cavity. As already discussed above, the width of the cavity resonance is reduced for smaller Drude dampening parameters, $\gamma$. For this reason, Fig. \ref{fig:spherical_cavity_benzene}a shows the 
absorption spectrum for the case with the true gold dampening, as well as 25\%, 10\% and 5\% of the dampening respectively. We mention in passing that in practice one could imagine realizing these lower losses using for example metals specifically engineered to show weaker losses~\cite{gjerding2017band}. Furthermore, one could also imagine exploring different cavity setups potentially leveraging the lower losses in dielectric nano-optical setups~\cite{kuznetsov2016optically}. Considering the wide range of available materials this design space becomes enormous~\cite{baranov2017all,svendsen2022computational}. As shown in the inset of Fig. \ref{fig:spherical_cavity_benzene}a, we find that reducing the losses results in a narrower cavity mode without a significant reduction in the overall cavity field strength. At around 25\% of the true dampening, we begin to observe clear indications of the two polariton peaks in the linear absorption spectrum. Further reducing the dampening we see clear strong coupling with a Rabi splitting of around $8 \mathrm{\,meV}$. This emphasizes the importance of the optical losses in reaching the strong coupling regime, and further highlights the significant strength of our method that we are able to study the effect of different cavity parameters from first principles via our combination of QEDFT and MQED.

Another way to engineer the coupling strength is by changing the emitter. There are two ways to do this, either by changing the number of emitters or by changing the emitter itself. To investigate the first option, we take the cavity with 25\% of the true gold losses, and compute the absorption spectrum for one, two and three benzene molecules, all of which we place in the center of the cavity. As shown in Fig. \ref{fig:spherical_cavity_benzene}b, we see a clear evolution of the Rabi splitting with the number of benzene molecules indicating the onset of collective strong coupling~\cite{ebbesen2016hybrid,genet2021inducing,sidler2022perspective,ruggenthaler2022understanding}. To perform this analysis, one needs to solve a coupled many electron, many photon problem and it again highlights the strength of the method.

\begin{figure*}[tb]
    \centering
    \includegraphics[width=\textwidth]{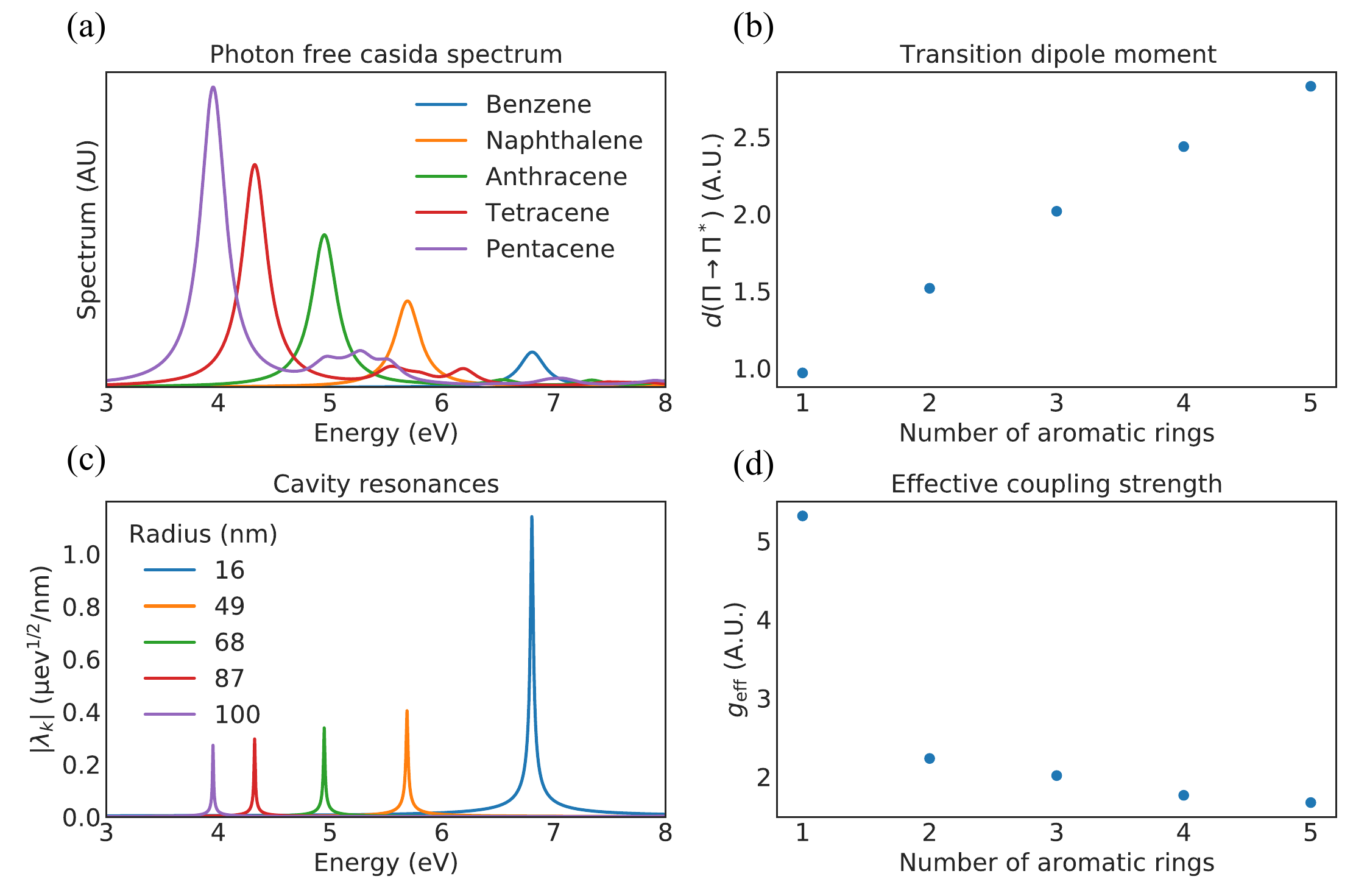}
    \caption{(a) The Casida spectrum without photon modes for different aromatic compounds. The spectra here are shown with an artificial broadening of 0.1361 eV because the photon-free calculations fail to naturally describe the linewidth of the transitions. (b) Transition dipole moment of the $\Pi\rightarrow\Pi^*$-transition as a number of aromatic rings. (c) Mode structure of the smallest cavities which host resonances aligned with the $\Pi\rightarrow\Pi^*$-transition of the different molecules. The mode structure is shown with a sampling density of $10$ points/meV. (d) Effective light-matter coupling strength as a function of the number of benzene rings in the aromatic molecule.}
    \label{fig:aromatic-compounds-screening}
\end{figure*}

To investigate the latter option, we consider longer aromatic compounds with $N$ aromatic rings; naphthalene $(N=2)$, anthracene $(N=3)$, tetracene $(N=4)$ and pentacene $(N=5)$. We first use the Casida method without photons (standard TDDFT) to characterize the spectral properties of the aromatic molecules. Fig.  \ref{fig:aromatic-compounds-screening}(a,b) shows respectively the spectrum and transition dipole moment of the $\Pi\rightarrow\Pi^*$-transition (HOMO-LUMO) as a function of the number of aromatic rings $N$. Note that in Fig.  \ref{fig:aromatic-compounds-screening}a the spectra are shown with the Octopus default artificial broadening of 0.1361 eV. This artificial broadening is necessary because unlike the combined MQED-QEDFT method, the standard photon-free TDDFT formulation fails to describe the linewidth of the transitions. We observe that with increasing number of aromatic rings $N$, the transition energy redshifts and the transition dipole moment increases linearly (Fig.  \ref{fig:aromatic-compounds-screening}b). The linearly increasing transition dipole moment would suggest that the light-matter coupling strength can be monotonically increased by simply using a larger aromatic molecule. However, because the transition energy also redshifts with increasing molecule length, the cavity has to be re-optimized to be resonant with the transition for each molecule, as shown in Fig.  \ref{fig:aromatic-compounds-screening}c. Specifically, focusing on modifications of the $R=16$ nm cavity, we find that this re-optimization of the cavity means that the cavity radius needs to be increased. This increase in radius leads to a reduced field concentration via an increased effective mode volume. This behavior highlights the important point that the light-matter coupling strength is a joint property of both the electronic system and the electromagnetic environment. A proper treatment of both is therefore essential for quantitative predictions.

We can characterize the intricate interplay between the electromagnetic environment and the electronic structure by looking at an effective coupling strength for the cavity modes which we define as~\cite{wang2021light},
\begin{align}
    g_\mathrm{eff} \propto \sqrt{\omega_c}\lambda_c |\boldsymbol d|.
\end{align}
Here the total cavity field strength parameter is defined as the coupling strength averaged across the cavity peak $p$, $\lambda_c = \sqrt{\int_p d\omega|\boldsymbol\lambda_d(\omega)|^2}$, and $\omega_c$ is the center frequency of the cavity mode. The $d$ subscript indicates that we take the field strength parameter for dipole orientation $\boldsymbol d/|\boldsymbol d|$. $g_\mathrm{eff}$ would thus be the true light-matter coupling if the total spectral weight was concentrated in a single mode. 
As shown in Fig.  \ref{fig:aromatic-compounds-screening}d,  we observe that the increase in transition dipole moment is counteracted by the reduced field concentration for the larger molecules, effectively resulting in a weaker light-matter coupling strength. This is in stark contrast to the intuitive argument based solely on the increased dipole moment of the longer molecules. 
It should be noted that $g_\mathrm{eff}$ is not a perfect measure of the light-matter coupling strength and it only gives a rough idea of the cavity re-optimization's effect. This is because the widths of the modes are not taken into account which, as we have seen in Fig. \ref{fig:spherical_cavity_benzene}a, is very important for the nature of the light-matter coupling.

Importantly, the coupling strength does not reach a significant fraction of the transitions frequencies for any of the cases considered above. This highlights that reaching the ultra- and deep strong coupling with single or few molecules is in general hard. For the coupling strengths we find in this paper, the results from the combined QEDFT-MQED method would therefore agree with those found using the method presented in Ref.~\cite{svendsen2021combining}. With the presented framework it would be possible to perform further engineering of the electromagnetic environment to increase the coupling. The application of the framework to general electromagnetic environments is discussed further below.

\subsection{Comment on Fabry Perot cavities}
A common example of a cavity in the literature for both theory and experiments is the layered Fabry Perot cavity (FPC). As discussed in Supplementary Note \ref{app:fabry-perot-cavity}, the FPC is a layered system and consequently its DGF is expanded in terms of in-plane plane waves, augmented by a function accounting for the reflection at the interfaces between the layers~\cite{chew1990waves}. Because the FPC only constrains the electromagnetic modes in one direction, it retains significant dispersion of the modes in plane. Consequently, the concentration of electromagnetic density of states is significantly less efficient than in the case of e.g. the spherical microcavity. This means that the resulting coupling strength is weaker and the FPC will therefore generally not be suited for single- or few emitter strong coupling~\cite{dutra1996spontaneous}. For this reason, we do not perform explicit QEDFT calculations for this cavity setup. However, the FPC can still be suited for collective strong coupling~\cite{ebbesenTilting} and coupling to extended systems~\cite{Gu_2021} where the extended modes of the electromagnetic environment can be sampled more effectively.

\subsection{Tool for cavity field parameters in simple cavities}
As a part of this work, we are making the code to generate the cavity field strengths available for everyone to use as part of the new \href{https://github.com/flickgroup/photonpilot}{PhotonPilot} tool. This tool currently allows the user to calculate cavity field strengths for the spherical and layered cavity setups, and we plan to expand its capabilities in the future.

\subsection{General electromagnetic environments}
We emphasize that the method we have presented here is general and applicable to any electromagnetic environment as long as the DGF can be determined. However, in  general electromagnetic environments with lower symmetry it is not possible to write down an analytical expression for the DGF. In such cases, the DGF must be constructed numerically from e.g. a mode expansion based on finite element simulations~\cite{chew1990waves,novotny2012principles}. We note in passing that in the general setting the Helmholtz equation is not a Hermitian operator. Special care is therefore needed when constructing the spectral representation of the DGF from the modes. One solution is to use the biorthonormal construction discussed in Ref. \cite{chen2010finite}. We envision that the method presented in this paper will eventually be integrated with the existing Maxwell solver in the Octopus code\cite{tancogne2020octopus,jestadt2019light}. Such an integration would allow for the treatment of general electromagnetic environments completely within Octopus.

\section{Conclusion}

In this paper, we have presented a methodology combining macroscopic quantum electrodynamics with quantum-electrodynamical density-functional theory which provides a fully \textit{ab initio} description of coupled quantum light-matter systems. Importantly, while we describe the electronic system at the DFT level, it is also possible to employ our approach to quantitatively describe the electromagnetic environment within standard few-level models of strongly coupled light-matter systems such as e.g. the Jaynes-Cummings model, the Rabi model, or the Travis-Cummings model.

To exemplify this approach, we have considered a benzene molecule strongly coupled to a metallic spherical cavity and investigated the impact of the both cavity radius and cavity loss on the nature of the light-matter coupling. We have further investigated the effect of adding more molecules and exchanging benzene with larger aromatic molecules. Together, these results highlight the intricate interplay between the electronic structure of the emitter and the environment in determining the nature of the light-matter coupling. Our work therefore illustrates the importance of having a proper description of both the electronic system and the electromagnetic environment for a proper description of quantum light-matter interactions. This work sets out the direction for more quantitative calculations in the future and also opens the possibility for the proper treatment of real experimental setups. We emphasize that the connection between the optical environment and the DGF is not limited to setups similar to cavities, but instead provides a general way to determine the electromagnetic spectral density of an arbitrary environment. In addition to the QED setup, our method therefore also provides a way to perform time-dependent density-functional theory in a lossy optical environment without the need for artificial spectral broadening.

Finally, we have provided an easy to use tool that everyone can use to generate cavity parameters for simple cavities such as the spherical microcavity or a layered cavity.

\begin{acknowledgement}

All calculations were performed using the computational facilities of the Flatiron Institute. The Flatiron Institute is a division of the Simons Foundation. We acknowledge support from the Max Planck-New York City Center for Non-Equilibrium Quantum Phenomena.
K.S.T. acknowledges funding from the European Research Council
(ERC) under the European Union’s Horizon 2020 research
and innovation program grant no. 773122 (LIMA). K.S.T. is a Villum
Investigator supported by VILLUM FONDEN (grant no.
37789).
J.F. acknowledges the support of the NSF Phase II CREST Center IDEALS with grant number EES-2112550).

\end{acknowledgement}

\begin{suppinfo}
The Supporting Information is available at \href{http://pubs.acs.org}{https://pubs.acs.org/doi/10.1021/acs.jctc.3c00967}.
\begin{itemize}
    \item Length-gauge Pauli Fierz Hamiltonian in terms of the emitter centered modes
    \item Dyadic Green's Function of the spherical microcavity
    \item Discretization of the cavity coupling parameters
    \item Computational details of the QEDFT calculations
    \item Connection to the standard light-matter coupling strengths used in quantum optics
    \item The Fabry Perot cavity
\end{itemize}

\end{suppinfo}

\bibliography{bib}

\end{document}